# Charge carrier mobility and nonproportionality of LaBr₃:Ce scintillators

I.V. Khodyuk, F.G.A. Quarati, M.S. Alekhin, and P. Dorenbos

*Luminescence Materials Research Group, Faculty of Applied Sciences,*
*Delft University of Technology, Mekelweg 15, Delft, 2629JB, The Netherlands*

The nonproportional response and related energy resolution of LaBr₃:Ce³⁺ scintillation crystals doped with different concentrations of cerium were studied between 80K and 450K. For Ce³⁺ concentration of 5% and 30%, LaBr₃ showed best proportionality and energy resolution at 80K. For LaBr₃:0.2%Ce the best energy resolution and the lowest degree of nonproportional response were instead observed around room temperature. The experimental results were analyzed in terms of charge carrier mobility and using theory of carrier transport in wide band gap semiconductors. We found that scattering of carriers by both lattice and impurity are the key processes determining the particular temperature dependence of carrier mobility and ultimately the scintillation nonproportionality. The calculated maximum of the LaBr₃:0.2%Ce carrier mobility corresponds well with the experimentally observed minima of its degree of nonproportionality, when assuming about 100ppm ionized impurity concentration.

## I Introduction

The dynamics of hot charge carriers created in the ionization track of ionizing particles is of interest in various disciplines of science. In a small cylindrical volume with radius r ~ 5 nm around the ionization track[1] schematically shown in Fig. 1 on the ps time scale[2] a very high ionization density n > 10²⁰ e-h/cm³ of free electrons and holes are created[3,4] that can cause secondary effects. For instance the energy density available is sufficient to displace atoms from their normal lattice positions thus creating radiation damage.[5] In tissue radiation damage may have severe health risks, and in dosimetry it can lead to underestimation of the total absorbed dose. Currently there are many investigations in utilizing carrier multiplication to develop better efficiency photo-voltaic cells. In inorganic scintillators, that is the topic of this work, the created free charge carriers need to escape the volume of high ionization density to be trapped by a luminescence center and recombine under emission of photons.[6]

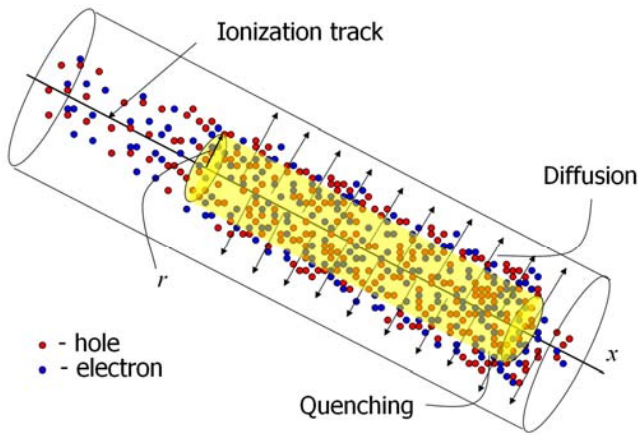

Fig. 1 (Color online) Sketch of an ionization track formed by a primary electron starting from the left creating free electrons and holes that diffuse radially away from the track. Radiationless carrier recombination occurs at the dense carrier concentration regions.

Scintillation crystals are widely used as spectroscopic detectors of ionizing radiation in nuclear science, space exploration, medical imaging, homeland security, etc. The important parameters for X- or $\gamma$-ray spectrometry are the total light output by the scintillator expressed in photons emitted per MeV of absorbed ionizing energy, decay time of the scintillation light flash, energy resolution for the detection of the ionizing particle and the detection efficiency. Taking into account all parameters one of the best inorganic scintillator commercially available today is $LaBr_3$:Ce. Concerning high light output and good energy resolution the rediscovered[7-9] $SrI_2$:Eu and recently discovered[10] $CsBa_2I_2$:Eu scintillators are very promising.

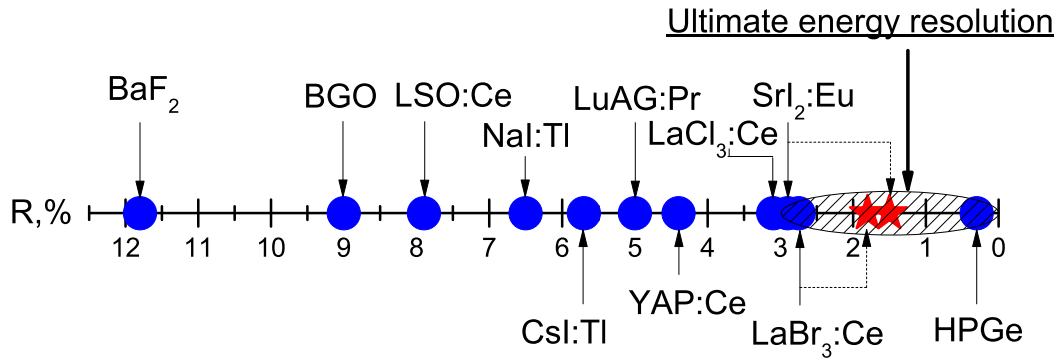

Fig. 2 (Color online) Energy resolution of inorganic scintillators and of a HPGe detector for the detection of 662 keV gamma photons. The energy resolution is defined as the full width at half maximum of total energy peak in scintillation pulse height spectra divided by the mean energy of that peak.

In spite of all efforts by the scintillation community the energy resolution of inorganic scintillators is still much larger than the fundamental limit dictated by photon statistics.[11] Figure 2 shows the energy resolution achieved by well-known scintillators for the detection of 662 keV gamma ray photons. The best resolution is for $LaBr_3$:Ce followed by $SrI_2$:Eu. The star symbols are the fundamental limit as dictated by photon statistics[12] for these two scintillators which demonstrates that there is still very significant improvement possible to well below 2%. For a solid state detector like high purity germanium (HPGe) photon statistics does not contribute and much better resolution down to 0.3% can be obtained. To decrease the energy resolution by almost a factor of two to 1.8% for $LaBr_3$ and to 1.5% for $SrI_2$ it is necessary to minimize all contributions other than photon statistics that influence energy resolution. The most essential contribution to be minimized is the contribution determined by nonproportionality.[13]

Nonproportionality is the nonlinear dependence of the total light output of the scintillator on the absorbed amount of ionization energy, i.e., the emitted number of photons/MeV at 10 keV is not necessarily the same as at 100 keV or at 1000 keV. This dependence is due to a scintillation efficiency that, in tern, depends on the density of the ionization track. The production of secondary electrons (i.e., Auger electrons, delta-rays etc.) during slowing down of the primary electron is a probabilistic process and may occur in different ways for the same absorbed energy. The dependence of the absolute light yield on the energy of secondary electrons and the probabilistic mechanism of their creation result in variability of the total number of photons produced inside the scintillator.[14] This process leads to broadening of the full-energy peak in the energy spectrum measured by a scintillation detector.

The nonproportionality of scintillators is attributed to radiationless recombination of electron-hole pairs with a recombination rate that increases with the ionization density.[2, 6, 15-19] This process together with an ionization density that changes along an electron track and with primary electron energy causes the deterioration of the energy resolution. To avoid the recombination losses, charge carriers should be effectively transferred from the primary track to luminescence centers. The faster the charge carriers escape the volume of high ionization density shown in Fig.1, in which quenching occurs, the higher the probability of converting carriers into optical photons. An important factor determining the rate at which carriers leave this volume is the carrier diffusion coefficient.[2, 6, 17] A high diffusion coefficient contributes to a more rapid transport of electrons, holes and excitons to regions further from the track where the radiationless recombination rate does not depend on ionization density.

In this paper the dependence of LaBr$_3$ nonproportionality on temperature and Ce$^{3+}$ concentration has been studied. For LaBr$_3$ with 0.2%, 5% and 30% of Ce$^{3+}$ the nonproportional response is determined at 80K, 300K and 450K and as a function of photon energy (photon-nPR) and as a function of electron energy (electron-nPR). Scintillation yield and energy resolution was measured in the energy range from 10.5 keV to 100 keV and at 662 keV. A specific model will be presented able to predict the electron-nPR results, and the degree of electron-nPR will be introduced and determined. Its dependence on temperature and concentration will be compared with our model estimate of the mobility for thermalized carriers in wide band gap semiconductors.

## II Experimental methods

To record scintillation pulse height spectra as a function of temperature, a LaBr$_3$:Ce sample was fixed at the bottom of a parabolic-like stainless steel cup covered with a reflective Al-foil, mounted onto the cold finger of a liquid nitrogen bath cryostat. The cup directs the scintillation light through a quartz window towards a photomultiplier tube (PMT) situated outside the cryostat chamber.[20] The Hamamatsu R6231-100 PMT at -680V bias voltage remained at room temperature and observes about 20% of the emitted scintillation light. To collect as much of the PMT output charge pulse as possible, the shaping time of an Ortec 672 spectroscopic amplifier was set at 10 µs. The temperature of the sample was controlled by two thermocouples attached to different parts of the sample holder. The yield of the scintillator will be expressed by the number of photoelectrons created in the PMT per MeV (N$_{phe}^{PMT}$/MeV) of absorbed gamma or X-ray photon energy. The energy resolution $R$ of a peak in the pulse height spectrum at energy $E$ is defined as the ratio of the full width at half maximum $\Delta E$ of that peak to the energy $E$, and it will be expressed as a percentage value.

To measure X-ray pulse height spectra at many finely spaced energy values between 10.5 keV and 100 keV, experiments were carried out at the X-1 beam line at the Hamburger Synchrotronstrahlungslabor (HASYLAB) synchrotron radiation facility in Hamburg, Germany. A highly monochromatic pencil X-ray beam in the energy range 10.5 – 100 keV was used as an excitation source. A tunable double Bragg reflection monochromator using a Si[511] set of silicon crystals providing an X-ray resolution of 1 eV at 10.5 keV rising to 20 eV at 100 keV was used to select the X-ray energies. A sketch of the experimental set-up can be found in Ref.[21, 22] The beam spot size was set by a pair of precision stepper-driven slits, positioned immediately in front of the cryostat chamber. For all measurements, a slit size of 50 × 50 µm$^2$ was used.

A dense sampling of data performed around the lanthanum K-electron binding energy $E_{KLa}$=38.925 keV was done in order to apply the K-dip spectroscopy method.[23] This method allows to derive the response of LaBr$_3$:Ce to photoelectrons down to energies as low as 100 eV. The method is briefly described as follows. An X-ray with energy $E_X$ that photoelectrically interacts with the lanthanum K-shell leads to the creation of a photoelectron with energy $E_e$ and a hole in the lanthanum K-shell,

$$E_e = E_X - E_{KLa} . \qquad (1)$$

The hole relaxes to the ground state with the emission of a cascade of secondary X-ray fluorescence photons and/or Auger electrons. The response of a scintillator is then equivalent to the sum of two main interaction products: 1) the K-shell photo electron response and 2) the response from the electrons emitted due to the sequence of processes following relaxation of the hole in the K-shell, the so-called K-cascade response. Our strategy is to employ X-ray energies just above $E_{KLa}$. The K-cascade response is assumed to be independent from the original X-ray energy. This response is found by tuning the X-ray energy to just above $E_{KLa}$. By subtracting the K-cascade response from the total X-ray response we are left with the response in photoelectrons from the K-shell photoelectron alone with energy $E_e$. The K-electron-nPR curve is then obtained from the number $N_{phe}^{PMT}$/MeV at the energy of the K-photoelectron divided by the number $N_{phe}^{PMT}$/MeV measured at 662 keV.

## III Results

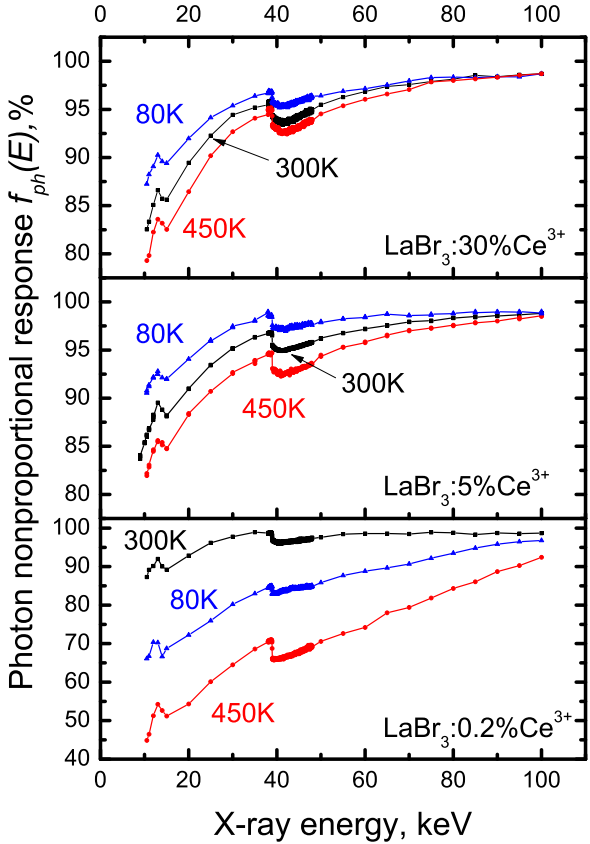

Fig. 3 (Color online) Photon nonproportional response of LaBr₃ doped with 0.2%, 5% and 30% $Ce^{3+}$ as a function of X-ray or gamma photon energy at 80K, 300K and 450K.

The photon nonproportional response (photon-nPR) written as $f_{ph}(E)$ is defined as the number of photoelectrons $N_{phe}^{PMT}$/MeV of absorbed energy observed at energy $E$ divided by the number $N_{phe}^{PMT}$/MeV observed at $E = 662$ keV energy. $f_{ph}(E)$ is expressed as a percentage value. For an ideal proportional scintillator it is 100% at all energies. Figure 3 shows $f_{ph}(E)$ for LaBr₃ doped with 0.2%, 5% and 30% $Ce^{3+}$ studied at 80K, 300K, and 450K. The shape of the $f_{ph}(E)$ curve depends not only on the temperature as was reported before,[16] but also on $Ce^{3+}$ concentration.

As a figure of merit the degree of photon-nPR $\sigma_{ph}$ will be used. It has been defined following ideas in[13, 24, 25]

$$\sigma_{ph} = \frac{1}{(E_{max} - E_{min})} \int_{E_{min}}^{E_{max}} \left| f_{ph}(E_{max}) - f_{ph}(E) \right| dE, \qquad (2)$$

where $E_{max}$ = 662 keV, $E_{min}$ = 10.5 keV, and $f_{ph}(E_{max})$ is set equal to 100%. $\sigma_{ph}$ for LaBr$_3$ at different temperatures and Ce$^{3+}$ concentrations obtained from the results in Fig. 3 are listed in Table I. For LaBr$_3$:5%Ce and LaBr$_3$:30%Ce $\sigma_{ph}$ increases with temperature. The behavior is different for LaBr$_3$:0.2%Ce where the lowest value for $\sigma_{ph}$ is observed at 300K. The smallest $\sigma_{ph}$ is measured for LaBr$_3$:5%Ce at 80K.

Table I. Degree (in %) of LaBr$_3$:Ce photon-nPR $\sigma_{ph}$ in the energy range from $E_{min}$ = 10.5 keV to $E_{max}$ = 662 keV.

| Ce$^{3+}$ concentration (%) | Temperature, K | | |
|---|---|---|---|
| | 80 | 300 | 450 |
| 0.2 | 3.31 | 0.95 | 6.98 |
| 5 | 0.78 | 1.07 | 1.43 |
| 30 | 1.09 | 1.22 | 1.37 |

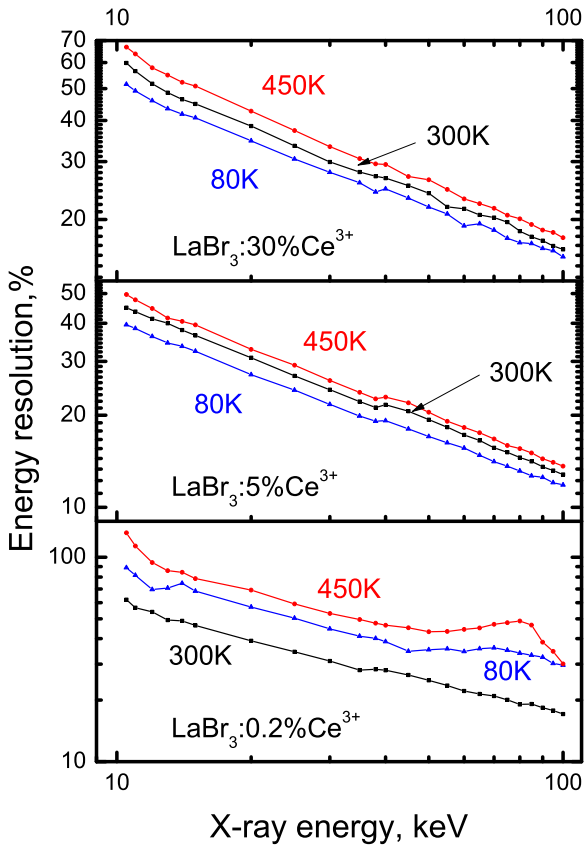

Fig. 4 (Color online) Energy resolution of LaBr$_3$ doped with 0.2%, 5% and 30% Ce$^{3+}$ as a function of X-ray energy at 80K, 300K, and 450K.

The energy resolution $R(E)$ of LaBr$_3$ doped with 0.2%, 5% and 30% Ce$^{3+}$ at 80K, 300K and 450K is presented in Fig. 4. The overall pattern is consistent with the pattern of $\sigma_{ph}$. At a given energy for both LaBr$_3$ doped with 5% and 30% Ce$^{3+}$ the best energy resolution is obtained at 80K and the worst at 450K. LaBr$_3$:0.2%Ce shows the best resolution at 300K where $\sigma_{ph}$ is minimal. Figure 4 shows that at 80K, the already outstanding room temperature energy resolution of LaBr$_3$ doped with 5 and 30% Ce$^{3+}$ can be improved even further. To confirm the dependence of $R$ on temperature and concentration, pulse height spectra were recorded using $^{137}$Cs 662 keV gamma radiation.

The energy resolution $\Delta E/E$ of a scintillator is determined by

$$\left(\frac{\Delta E}{E}\right)^2 = R^2 = R_M^2(T) + R_{sc}^2(T) = (2.35)^2 \frac{1+v(M)}{N_{phe}^{PMT}(T)} + R_{sc}^2(T), \qquad (3)$$

where $v(M)$ is the variance in the PMT gain, $N_{phe}^{PMT}$ is the number of photoelectrons that are produced by the interaction of scintillation photons with the PMT photocathode and are multiplied on the first dynode,[26, 27] and $R_{sc}$ is given by

$$R_{sc}^2(T) = R_{nPR}^2(T) + R_{tr}^2(T) + R_{inh}^2(T), \qquad (4)$$

where $R_{nPR}(T)$ is a contribution from nonproportionality, $R_{tr}(T)$ is the so-called transport resolution, and $R_{inh}(T)$ is a contribution from inhomogeneity of the scintillation crystal. It is assumed that all contributions are independent from each other.

To measure the temperature dependence of the LaBr$_3$:Ce energy resolution the parabolic-like cup covered with reflective Al foil was used. This configuration of the experimental set-up results in the collection of about 20% of the emitted scintillation photons, increasing importance of the statistical contribution $R_M(T)$.

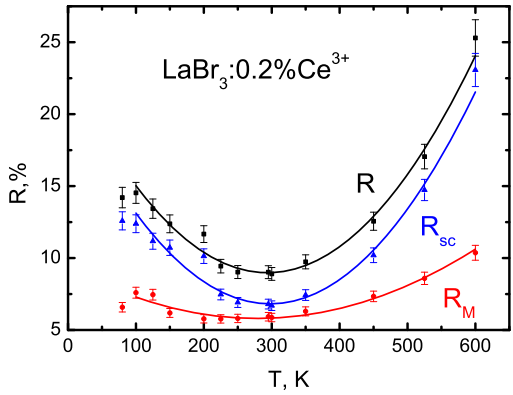

Fig. 5 (Color online) The separate contributions to the total energy resolution of LaBr$_3$:0.2% Ce at 662 keV as a function of temperature.

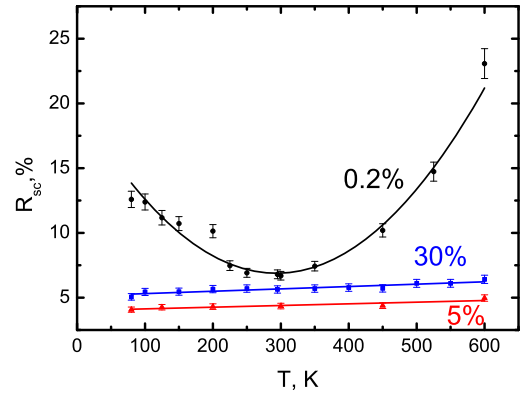

Fig. 6 (Color online) The $R_{sc}$ contribution to the energy resolution at 662 keV of 0.2%, 5% and 30% Ce-doped LaBr$_3$ as a function of temperature.

Figure 5 shows the measured $R(T)$, $R_M(T)$ calculated from the measured $N_{phe}^{PMT}$, and $R_{sc}(T)$ obtained with Eq. (4) for LaBr$_3$:0.2%Ce. The parabolas through the data are drawn to guide the eye. $R_M(T)$ is small and $R(T)$ is almost entirely determined by $R_{sc}(T)$. The resolution is lowest at room temperature. This pattern is consistent with the pattern of $\sigma_{ph}$ in Table I where a larger $\sigma_{ph}$ results in poorer energy resolution which confirms a relationship between energy resolution and nonproportionality.

The contribution $R_{sc}(T)$ to the energy resolution at 662 keV is shown in Fig. 6 for 0.2%, 5% and 30% Ce-doped LaBr$_3$. LaBr$_3$:0.2%Ce shows a minimum at room temperature. In contrast, LaBr$_3$ with 5% and 30% Ce$^{3+}$ exhibits a linear decrease of the $R_{sc}(T)$ with decreasing temperature. Lower values of $R_{sc}(T)$ correlate with lower values of $\sigma_{ph}$.

Using K-dip spectroscopy we derived the K-photoelectron-nPR curves $f_e(E)$ for LaBr$_3$ doped with 0.2%, 5% and 30% Ce$^{3+}$ at 80K, 300K and 450K which are shown in Fig. 7.

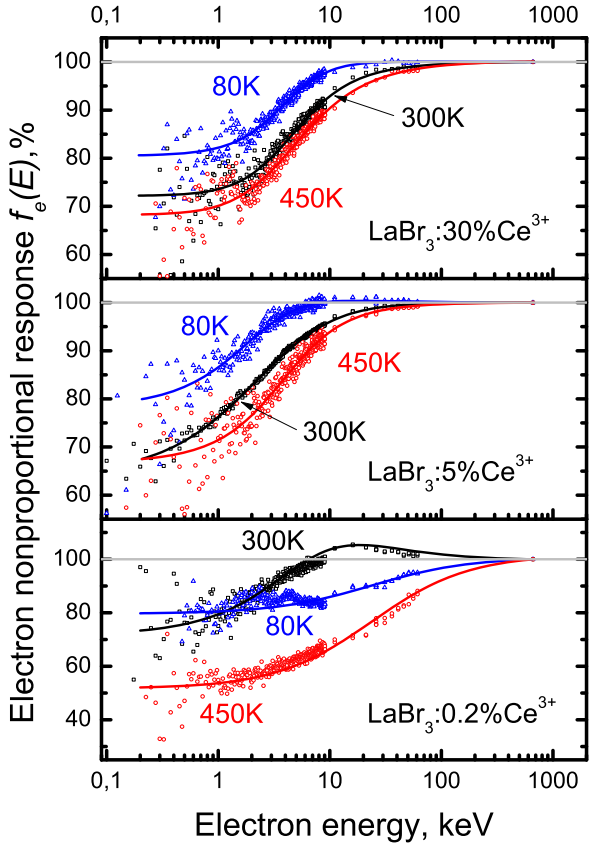

Fig. 7 (Color online) K-photoelectron nonproportional response of LaBr$_3$ doped with 0.2%, 5% and 30% Ce$^{3+}$ as a function of X-ray or gamma photon energy at 80K, 300K and 450K.

The $f_e(E)$ data in Fig. 7 are quite scattered, especially below 1 keV.[28] To represent the data with a smoothly varying curve an approach similar to the algorithms used in[15, 29, 30] was employed. Using the nonrelativistic Bethe equation the rate of energy loss by the primary electron or stopping power[31] can be written as

$$-\frac{dE}{dx} = \frac{2\pi e^4 \left(4\pi\varepsilon_0\right)\rho_e}{E}\ln\left\{1.164\left[E+0.81I\right]/I\right\}, \qquad (5)$$

where $e$ is the elementary charge, $\varepsilon_0$ is the vacuum dielectric permittivity, $\rho_e$ is the electron density in the scintillator, $E$ is the energy of the ionization track creating electron, and $I$ is the average ionization energy of the Hydrogen atom. Assuming cylindrical shape of high ionization density volume[1] along the track of the primary energetic electron as shown in Fig.1, the concentration of the ionized charge carriers $n(x)$ is given by

$$n\left(x\right) = \frac{1}{\pi r^2 E_{e-h}}\left(-\frac{dE}{dx}\right), \qquad (6)$$

where $r$ is the radius of the high ionization density volume shown in Fig. 1 and $E_{e-h}$ is the average energy required to create a free electron- free hole pair in the scintillator.[5, 32]

Finally, following the ideas in Ref.[15, 30] $f_e(E)$ can be represented by

$$f_e(E) = \left[ A1 + \frac{A2 + A3 \cdot n(x)}{A4 + A5 \cdot n(x) + A6 \cdot n^2(x)} \right] \times 100\%, \qquad (7)$$

where A1 to A6 are independent fitting parameters. The results are shown by the solid curves in Figure 7, and will be used to calculate the degree of electron-nPR $\sigma_e$.

$\sigma_e$ is defined analogous to the degree of photon-nPR and is determined using Eq. (2) by integrating over the energy range from $E_{min} = 0.2$ keV to $E_{max} = 662$ keV. For a perfectly proportional scintillator the value of $\sigma_e$ is zero, and the scintillator with a lower value of $\sigma_e$ is considered to be more proportional.

Table II. Degree (in %) of the LaBr$_3$:Ce electron-nPR $\sigma_e$ in the energy range from $E_{min} = 0.2$ keV to $E_{max} = 662$ keV.

| Ce$^{3+}$ concentration (%) | Temperature, K | | |
|---|---|---|---|
| | 80 | 300 | 450 |
| 0.2 | 1.80 | 0.93 | 4.28 |
| 5 | 0.12 | 0.29 | 0.45 |
| 30 | 0.16 | 0.52 | 0.68 |

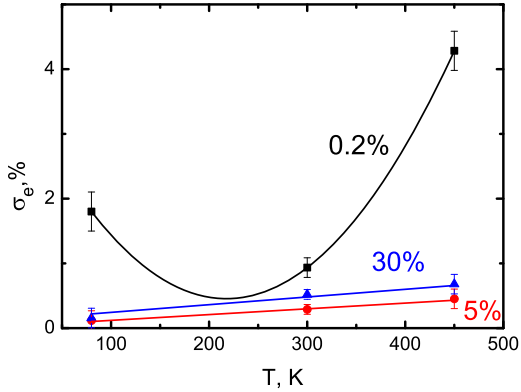

Fig. 8 (Color online) Degree of LaBr$_3$ electron-nPR $\sigma_e$ versus temperature and Ce$^{3+}$ concentration.

$\sigma_e$ versus $T$ and Ce$^{3+}$ concentration is shown in Fig. 8 and in Table II. It behaves similar to $\sigma_{ph}$. The only difference is that $\sigma_e$ of LaBr$_3$:0.2%Ce at 300K shows a higher value of 0.93% compared to 0.29% for LaBr$_3$:5%Ce and 0.52% for LaBr$_3$:30%Ce. Linear extrapolation of $\sigma_e$ for LaBr$_3$:5%Ce and LaBr$_3$:30%Ce suggests that $\sigma_e$ for both concentrations reach zero at a temperature close to the absolute zero. This means, that an almost perfect proportional response would be obtained for 5% and for 30% Ce-doped LaBr$_3$ crystals.

**IV Discussion**

Using synchrotron irradiation the photon-nPR $f_{ph}(E)$ and energy resolution $R$ of LaBr$_3$:Ce scintillation crystals doped with 0.2%, 5% and 30% of Ce$^{3+}$ were studied at 80K, 300K and 450K. Results of these experiments were shown in Figs. 3 and 4, and in Table I. $f_{ph}(E)$ and $\sigma_{ph}$ are characteristics of the gamma photon response of a scintillator, however, the response to energetic electrons is more fundamental. If $f_e(E)$ is known and when the process of ionization track creation can be simulated, the shape of $f_{ph}(E)$ over the entire energy range can be calculated[33, 34] by Monte-Carlo techniques. The actual value of $f_{ph}(E)$ at energy $E$ is then a weighted average of several values of $f_e(E)$ at lower energies.[14] Using the electron-nPR function $f_e(E)$ then provides a better starting point to understand nonproportionality then using the photon-nPR function.[14, 35] Using the K-dip spectroscopy method $f_e(E)$ shown in Fig. 7 was derived from the $f_{ph}(E)$ and Table II was calculated using Eq. (2) and Eq. (7). Figure 8 shows $\sigma_e$ of LaBr$_3$ versus temperature and Ce$^{3+}$ concentration, and this figure is the most important outcome of the performed experiments and calculations. In the following discussion we will concentrate on better understanding of the results in Fig. 8 by using ideas on carrier mobility from semiconductor physics and apply them to the processes that occur inside the ionization track in scintillators.

There are several models proposed in the recent literature to explain the origin of nonproportionality.[2, 6, 15, 25, 28, 36] It is attributed to radiationless electron-hole pair recombination in the regions of a high concentration $n(x)$ of charge carriers along the ionization track as shown in Fig. 1. According to Eq. (5) and Eq. (6) $n(x)$ increases with smaller energy $E$ of the track creating primary electron.[31] This leads to a larger radiationless electron hole recombination rate which forms the basis of increasing nonproportionality with smaller gamma or X-ray photon or primary electron energy.

An overview of the current models on nonproportionality was presented by Moses *et al.*[18] The basis of all those models is the competition between two opposing processes shown in Fig. 1: 1) quenching due to radiationless electron hole recombination inside the volume of high ionization density along the track, and 2) diffusion of the charge carriers from the point of creation towards a volume of lower ionization density. The faster the charge carriers escape the volume of high ionization density in which quenching occurs and reach luminescence centers, the higher the probability of converting the energy of the carriers into optical photons. An important factor determining the rate at which carriers leave this volume is the carrier diffusion coefficient.[2, 6, 17] Another very important parameter is concentration of luminescence or trapping centers inside the high ionization density volume. At high concentration of Ce$^{3+}$ in LaBr$_3$ essential part of the charge carriers can be promptly removed from the diffusion-quenching process. According to Bizarri and Dorenbos[37] carriers can be sequentially captured by Ce$^{3+}$ or form self-trapped excitons (STE) which transfer their energy to Ce$^{3+}$ through thermally activated migration or directly. These effects can lead to a significant difference of quenching probability at low Ce concentration 0.2% and at high concentrations 5% and 30%.

First let us consider processes which apply to all concentrations of luminescence centers. A high diffusion coefficient contributes to a more rapid transport of electrons, holes and excitons to regions away from the track where the radiationless recombination rate does not depend on concentration $n(x)$. Three different orders of radiationless recombination or quenching processes are distinguished in

$$\left.\frac{\partial n(r,t)}{\partial t}\right|_{quenching} = -k_i \cdot n^i(r,t), \qquad (8)$$

where $k_i(t)$ is the quenching rate and $i$ is the order of the quenching process. First order quenching includes: radiationless decay of the excited luminescence centre due to emission of phonons and plasmons, radiationless recombination of free excitons and STEs.[38] Second order quenching includes bimolecular quenching due to dipole-dipole Förster transfer.[3] And third order quenching includes exciton-exciton annihilation due to Auger-like processes.[2, 6] These different quenching processes are also at the bases[15] of the phenomenological Eq. (7) that determines the shape of $f_e(E)$.

The diffusion equation[39] can be written as

$$\left.\frac{\partial n(r,t)}{\partial t}\right|_{diffusion} = \nabla\Big[D(n)\nabla n(r,t)\Big], \qquad\qquad (9)$$

where $n$ is the concentration of charge carriers, $r$ is the radial coordinate perpendicular to the ionization track as shown in Fig.1, $t$ is the time and $D$ the diffusion coefficient. Assuming that in the high ionization density volume $D(n)$ is constant,[40] the Einstein relation

$$D = \mu(T) \cdot kT \qquad (10)$$

applies and one obtains

$$\left.\frac{\partial n(r,t)}{\partial t}\right|_{diffusion} = \mu(T) \cdot kT \cdot \nabla^2 n(r,t), \qquad (11)$$

where $\mu(T)$ is the mobility of the charge carriers, $k$ is the Boltzmann constant and $T$ is the effective temperature of the charge carriers. Based on Eq. (9), the transport of charge carriers becomes faster when carrier mobility and temperature increases.

The minimum in $\sigma_e$ for LaBr$_3$:0.2% Ce$^{3+}$ in Fig. 8 at room temperature suggests a minimum in the loss processes at room temperature that within the above theory should correspond with a maximum in charge carrier mobility. According to theory of charge carrier transport in wide band gap semiconductors, mobility indeed strongly depends on temperature.[40, 41] Here we will employ that theory in order to understand the results for LaBr$_3$:0.2% Ce$^{3+}$ in Fig. 8. The theory is for thermalized charge carriers and we therefore assume that all charge carriers are thermalized instantly[42, 43] after creation in the ionization track.

An increase of carrier mobility with temperature decrease is due to a reduced phonon interaction rate. Emission of optical phonons is the main mechanism responsible for carrier scattering by the lattice. LaBr$_3$ does not show any piezoelectric properties. That means that piezoelectric mode scattering caused by the electric field associated with acoustical phonons can also be ignored in our calculations. Lattice scattering due to optical phonons is independent on the carrier concentration.[40]

The main lattice scattering mechanism is due to the interaction of carriers with the longitudinal-optical phonons. According to Ref.[40] the optical Hall lattice mobility can be calculated from

$$\mu_{L_{opt}} = \frac{e}{2\alpha\omega_0 m^*}\left(\exp\left(\frac{\hbar\omega_0}{kT}\right)-1\right) \qquad (12)$$

where $\alpha$ is the polaron coupling constant given by

$$\alpha = \left(\frac{1}{\varepsilon_\infty} - \frac{1}{\varepsilon}\right)\sqrt{\frac{m^* E_H}{m_e \hbar\omega_0}} \qquad (13)$$

For LaBr$_3$ the high frequency and the static dielectric constants are $\varepsilon_\infty \approx 5$ and $\varepsilon \approx 10$, respectively;[44] $E_H = 13.595\,eV$ is the first ionization energy of the hydrogen atom; $\frac{m^*}{m_e} = 1.323$ is the effective electron mass divided by the electron mass;[25] $\hbar\omega_0 = 23.7\,meV$ is the energy of the longitudinal-optical phonon in LaBr$_3$.[44]

An increase of carrier mobility with temperature *increase* can be caused only by the ionized impurity scattering,[41] which according to Ziman[41] is given by

$$\mu_i = \frac{2^{7/2}(\varepsilon\varepsilon_0)^2(kT)^{3/2}}{\pi^{3/2}z^2 e^3(m^*)^{1/2}N_i} \cdot F(3kT) \qquad (14)$$

where $z$ is the effective charge of the impurity with concentration $N_i$, $\varepsilon_0$ is the vacuum permittivity and $F(3kT)$ is the averaged Coulomb screening factor[41]

$$F(3kT) = \ln\left(1 + \frac{8m* \cdot 3kT}{q^2\hbar^2}\right) - \left(1 + \frac{q^2\hbar^2}{8m* \cdot 3kT}\right)^{-1}, \qquad (15)$$

where

$$q^2 = \frac{4\pi e^2}{\varepsilon\varepsilon_0 kT} n\left(2 - \frac{n}{N_i}\right). \qquad (16)$$

For our range of temperatures $F(3kT) \approx 1$.

The overall mobility $\mu(T)$ can be obtained from

$$\mu(T) = \left(\frac{1}{\mu_L(T)} + \frac{1}{\mu_i(T)}\right)^{-1}. \qquad (17)$$

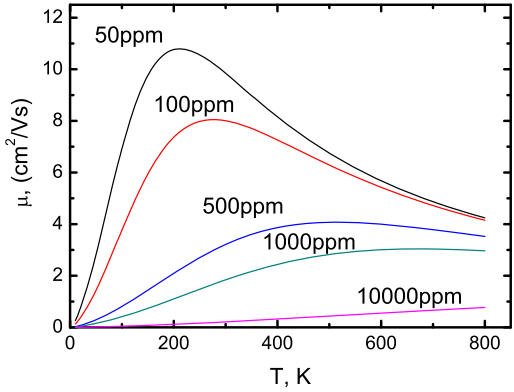

Fig. 9 (Color online) Calculated mobility of charge carriers in LaBr$_3$ versus temperature and ionized impurity concentration.

Figure 9 shows the mobility calculated with Eq. (17) for different concentrations of ionized impurity scattering centers with z = 1 in LaBr$_3$. At impurity concentration of 100ppm the maximum of the carrier mobility is slightly below room temperature. Therefore with our model an impurity concentration of 100ppm is needed to match well with the minimum of $R_{sc}(T)$ of LaBr$_3$:0.2%Ce at 300K in Fig. 5 and of $\sigma_e$ in Fig. 8.

Equations (12) to (16) pertain to a given density of carriers in the conduction or valence band. The calculations do not incorporate any carrier trapping[37] and also it was assumed that all charge carriers are thermalized instantly[42, 43] after creation in the ionization track. However recent theoretical studies[28, 45] suggest that also non-thermalized carriers play an important role in carrier and phonon transport in scintillators. One should therefore interpret the results in Fig. 9 as qualitative.

Lattice and impurity scattering mechanisms are expected to be more important at low Ce$^{3+}$ concentration due to the longer distance carriers need to travel before they can reach Ce$^{3+}$ where they can recombine radiatively. The concentration of Ce$^{3+}$ in LaBr$_3$:0.2%Ce is $4.2\cdot10^{18}$ $cm^{-3}$. At 5 and 30% Ce$^{3+}$ concentration the carrier density $n(x)$ is $1.05\cdot10^{20}$ $cm^{-3}$ and $6.3\cdot10^{20}$ $cm^{-3}$ which is of the same order of magnitude as the concentration of recombination centers, and a high mobility of charge carriers needed to escape the dense ionization region becomes of less importance. Carriers can be trapped instantly after ionization and the trapping rate by Ce$^{3+}$ starts to dominate over the quenching rate and the escape rate. This can explain the better $\sigma_e$ shown in Fig. 8 for 5 and 30% Ce concentration. What then still needs to be explained is the temperature dependence of $\sigma_e$ for 5 and 30% of Ce.

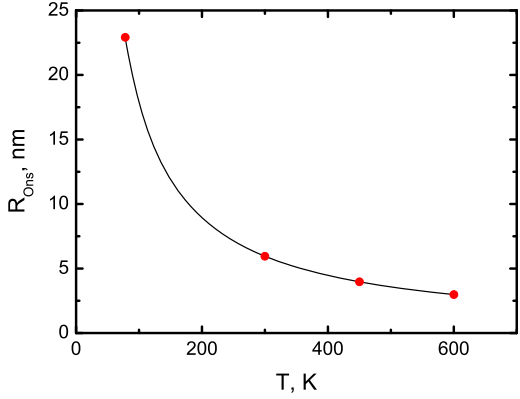

Fig. 10 (Color online) Onsager radius of charge carrier capture in LaBr$_3$ versus temperature.

According to the model of Bizarri[37] there are two routes for carriers to be trapped by Ce$^{3+}$ in LaBr$_3$. The first is the sequentially capture of first the hole by Ce$^{3+}$ with formation of Ce$^{4+}$ followed by capture of the electron, and the second through formation of STEs which transfer their energy to Ce$^{3+}$ though thermally activated migration or directly. Formation of STEs remains an important intermediate mechanism even at high Ce concentration, and especially at low temperatures. It means that quenching inside the high ionization density volume around the track can be reduced by more efficient formation of STEs. Then according to the Onsager model[46] and following ideas by Payne *et al.*[47] one may express the Onsager radius ($R_{Ons}$) of exciton formation as

$$\frac{e^2}{4\pi\varepsilon\varepsilon_0 R_{Ons}} = kT \; . \qquad (18)$$

Figure 10 shows the temperature dependence of $R_{Ons}$ for LaBr$_3$. Rapid and efficient formation of STEs and transfer of their energy to the luminescence centers removes carriers from the diffusion-quenching process shown in Fig. 1. Such mechanism will lower $\sigma_e$ and $R_{sc}(T)$ when temperature decreases, which then may explain the observations in Figs. 6 and 8.

## V Conclusion

The shape of the photon- and electron-nPR curves of LaBr$_3$:Ce depends on temperature. For 5% and 30% Ce$^{3+}$ concentration, LaBr$_3$ shows better proportionality and energy resolution when temperature decreases. This improvement means that at a low temperature even better energy resolution can be achieved with a LaBr$_3$ scintillation detector compared to the already outstanding 2.75% measured at room temperature.

The temperature dependence of the photon- and electron-nPRs of LaBr$_3$:0.2% Ce is different. The most proportional response was measured at 300K. At 80K and 450K the photon- and electron-nPR curves deviate strongly from the linear response. This leads to a significant deterioration of the energy resolution both at 80K and 450K.

Despite the limitations of the theoretical model that was used, the obtained results suggest that a significant factor determining the nonproportionality of LaBr$_3$:0.2%Ce is the mobility of charge carriers. The higher the carrier mobility and diffusion coefficient the lower the degree of electron-nPR, which leads to improved energy resolution. Semiconductor detectors based on HPGe with excellent energy resolution of 0.3% besides different statistics have a much higher mobility of charge carriers ~40000 $cm^2$/Vs compared to ~8 $cm^2$/Vs calculated for $LaBr_3$:0.2%Ce with 100ppm ionized impurity concentration. For 5% and 30% concentrations direct trapping by the recombination centers starts do dominate and a high mobility of charge carriers becomes of less importance.

Summarizing the results of the performed measurements and calculations and bearing in mind that carrier mobility in semiconductor detectors is high, we conclude that the "ultimate energy resolution" should be sought in scintillation materials with high carrier mobility and high charge carrier capture efficiency.

**Acknowledgments**

The research leading to these results has received funding from the Netherlands Technology Foundation (STW), Saint Gobain, crystals and detectors division, Nemours, France, and by the European Community's Seventh Framework Programme (FP7/2007-2013) under grant agreement n° 226716. We thank the scientists and technicians of the X-1 beamline at the Hamburger Synhrotronstrahlungslabor (HASY-LAB) synchrotron radiation facilities for their assistance. The authors want to acknowledge Conny Hansson, Johannes van der Biezen and Alan Owens from the European Space Agency (ESTEC) for their assistance with the experiment and sharing some of the beam time at X-1.